\documentclass{sigchi}

\CopyrightYear{2020}
\setcopyright{acmlicensed}
\doi{https://doi.org/10.1145/3313831.3376776}
\isbn{978-1-4503-6708-0/20/04}
\conferenceinfo{CHI'20,}{April  25--30, 2020, Honolulu, HI, USA}
\acmPrice{\$15.00}

\usepackage{balance}       
\usepackage{graphics}      
\usepackage[T1]{fontenc}   
\usepackage{txfonts}
\usepackage{mathptmx}
\usepackage[pdflang={en-US},pdftex]{hyperref}
\usepackage{color}
\usepackage{booktabs}
\usepackage{textcomp}
\usepackage{subfigure}
\usepackage{enumitem}
\usepackage{colortbl}
\definecolor{Gray}{gray}{0.9}
\usepackage{float}

\usepackage{microtype}        
\usepackage{ccicons}          

\usepackage{todonotes}

\def\plaintitle{Pedagogical Agents for Fostering \\Question-Asking Skills in Children}
\def\plainauthor{First Author, Second Author, Third Author,
  Fourth Author, Fifth Author, Sixth Author}

\def\plainkeywords{pedagogical agents; question-asking; educational application; divergent vs convergent thinking; epistemic curiosity}

\makeatletter
\def\url@leostyle{%
  \@ifundefined{selectfont}{
    \def\UrlFont{\sf}
  }{
    \def\UrlFont{\small\bf\ttfamily}
  }}
\makeatother
\def\pprw{8.5in}
\def\pprh{11in}

\definecolor{linkColor}{RGB}{6,125,233}
\hypersetup{%
  pdftitle={\plaintitle},
  pdfauthor={\plainauthor},
  pdfkeywords={\plainkeywords},
  pdfdisplaydoctitle=true, 
  bookmarksnumbered,
  pdfstartview={FitH},
  colorlinks,
  citecolor=black,
  filecolor=black,
  linkcolor=black,
  urlcolor=linkColor,
  breaklinks=true,
  hypertexnames=false
}

\begin{document}

\title{\plaintitle}

\numberofauthors{5}
\author{%
  \alignauthor{Mehdi Alaimi\\
    \affaddr{University of Bordeaux}\\
    \affaddr{Talence, France}}\\
    \email{\mbox{mehdi.alaimi@gmail.com}}\\
  \alignauthor{Edith Law\\
    \affaddr{University of Waterloo}\\
    \affaddr{Waterloo, ON, Canada}}\\
    \email{edith.law@uwaterloo.ca}\\
  \alignauthor{Kevin Daniel Pantasdo\\
    \affaddr{University of Waterloo}\\
    \affaddr{Waterloo, ON, Canada}}\\
    \email{kevin.pantasdo@edu.uwaterloo.ca}\\
  \alignauthor{Pierre-Yves Oudeyer\\
    \affaddr{Inria Bordeaux}\\
    \affaddr{Talence, France}}\\
    \email{pierre-yves.oudeyer@inria.fr}\\
  \alignauthor{H\'el\`ene Sauzeon\\
    \affaddr{Inria Bordeaux}\\
    \affaddr{Talence, France}}\\
    \email{helene.sauzeon@inria.fr}\\
}

\maketitle

\begin{abstract}
Question asking is an important tool for constructing academic knowledge, and a self-reinforcing driver of curiosity.   However, research has found that question asking is infrequent in the classroom and children's questions are often superficial, lacking deep reasoning.  In this work, we developed a pedagogical agent that encourages children to ask divergent-thinking questions, a more complex form of questions that is associated with curiosity. We conducted a study with 95 fifth grade students, who interacted with an agent that encourages either convergent-thinking or divergent-thinking questions. Results showed that both interventions increased the number of divergent-thinking questions and the fluency of question asking, while they did not significantly alter children's perception of curiosity despite their high intrinsic motivation scores.  In addition, children's curiosity trait has a mediating effect on question asking under the divergent-thinking agent, suggesting that question-asking interventions must be personalized to each student based on their tendency to be curious.
\end{abstract}

\ccsdesc[500]{Human-centered computing~Empirical studies in HCI}
\ccsdesc[500]{Applied computing~Interactive learning environments}

\keywords{\plainkeywords}

\printccsdesc

\section{Introduction}

A key challenge for 21st-century schools is the need to serve diverse students with varied abilities and motivations for learning \cite{durlak2011impact}. Active learning, a form of learning where students take initiative in the learning process, has been shown to help students develop their full potential by providing developmentally appropriate and individually tailored learning opportunities \cite{hohmann1995educating}. One way that children can take more initiative in their own learning is by asking question.  


Besides providing information, question-asking offers many other benefits.  On lecture comprehension tests, university students who generated their own questions got better scores than students who used questions from someone else \cite{king_autonomy_1994}.  Likewise, Davey and McBride \cite{Davey1986}  found that sixth grade students who were asked to create two good questions for each passage of the text performed better on comprehension tests than students who were asked to merely read.  The act of composing a question seems to focus students' attention on the most important/relevant information in the content \cite{brown_instructing_1984}.

Different types of questions offer different benefits.  For example, Gallagher and Ascher's hierarchical taxonomy \cite{gallagher1963preliminary} classified questions as low vs high level.  Low-level questions are surface-level, memory-based questions that ask students to name, define or identify (e.g. ``Who is the main character?''), or convergent-thinking questions that ask students to relate ideas by comparing, contrasting or explaining (e.g. ``Why was the character doing this at the beginning of the story?'').  High-level questions are deeper questions that involve divergent-thinking; they require responses that offer a new perspective on a given topic, asking students to predict, infer, hypothesize, reconstruct information or questions that incorporate new knowledge (e.g., ``What could happen if the main character did this instead of that?") and make subjective or moral judgements (e.g. ``What is your opinion about this?'').

Research has shown that there is an intimate relationship between epistemic curiosity and divergent-thinking question-asking.  Curiosity arises when one becomes aware of a knowledge gap; this awareness can lead one to ask questions in order to obtain the missing information \cite{ram_theory_1991}.   Children who are curious by trait have been shown to have better question-asking abilities \cite{jirout_childrens_2011}.  
The premise of our work is that question-asking is a skill beneficial to learning and the development of curiosity, that can be both taught and practiced.

In this work, we introduce a pedagogical agent designed to encourage students to ask divergent-thinking questions about pieces of text.  We conducted an experiment with 95 fifth grade students at an elementary school in France, who interacted with an agent that encouraged either convergent-thinking or divergent-thinking questions.  Results show that our experimental manipulation influenced the type and quantity of questions asked by children.  Furthermore, children found this question-asking learning exercise to be enjoyable and motivating.  Our work contributes both an educational platform for promoting question-asking skills, and insights into effective  technology-mediated interventions on question-asking.



\section{Related Work}

\subsection{Question Generation in Classroom}

Contrary to the popular belief that young children are avid questioners, research has shown that questions in classroom are not very frequent and are often low-level questions that do not necessitate deep reasoning \cite{graesser1994question,humphries2015beyond}. 

Low frequency of question asking can be attributed to three curiosity-based explanations. First, children may not be interested in asking questions; they are not motivated/curious because they can't identify their own knowledge gaps \cite{graesser1994question}. Second, social influences, from peers or teachers, can alter children's perception of curiosity, which in turn, fosters or inhibits their question asking behavior \cite{engel2011children}.  Post and Walma van der Molen \cite{post_development_2019}, for example, found that the fear of classmates' negative judgement has a detrimental effect on curiosity, causing children to have a negative opinion about asking question in a classroom (e.g. people who ask question are stupid).  Even the arrangement of the classroom, such as the positions of the tables, can have an impact on question generation; for instance, when children are assigned to sit in a semi-circle, they are more inclined to ask questions \cite{marx_effects_1999}. Most importantly, children may not know {\it how} to ask questions \cite{humphries2015beyond,graesser1994question}. During a think-aloud session, Humphries and Ness \cite{humphries2015beyond} measured the quantity and quality of questions generated by 5th grade students. The children had to read a piece of text and ask as many questions as they could while reading. They noticed children mainly ask questions using question starters, such as ``who'', ``what'', ``when'', ``where'', and that 93\% of their questions were low-level ones (memory-based or convergent-thinking questions). They concluded that children did not have the tools to help them construct higher-level thinking questions, such as ``What's the difference between ... ?'', ``What is your belief about ...?'', ``How do you know ...?'', etc \cite{humphries2015beyond,king_effects_1989,raphael_increasing_1985}.

In their review, Rosenshine, Meister and Chapman \cite{rosenshine_teaching_1996} compiled intervention studies that attempt to improve students' understanding of textual information by asking them to generate questions.  They grouped the studies according to the procedural prompts used to help children generate questions; these prompts include, for example, signal words for starting questions (e.g., who, what, where, when, why, how), requests for specific types of questions, or examples of questions. Overall, providing  signal words was the most effective way to improve the generation of questions from students. Teaching students to generate more questions was also found to enhance their comprehension. Ness \cite{ness_simple_2017} demonstrated to elementary school children how she generates simple and high-level questions from songs, and asked them to do the same on other songs. She noticed that, over time, the quality and quantity of the questions that the children ask improved.  These results suggest that it is possible to improve the question-asking skills of children by giving them tools they can reuse in classroom or daily life.

\subsection{Taxonomy of Questions}

Prior work has proposed many different frameworks for classifying questions.  Methods used in machine learning domains~\cite{collins2005discriminative,heilman2010good}, such as automatic question generation systems, see questions as different syntactic transformations (e.g subject-auxiliary inversion) of declarative sentences.   Questions are sorted by their length, meaning or complexity. These classification schemes are complex, and the groupings can be subjective and highly dependent on the interpretation of the annotator.

Other classification schemes, collectively known as QAR, consider the relationship between a question and its answer.  Raphael and Pearson \cite{raphael_increasing_1985} suggests that questions can be classified into three categories, depending on whether the answer is i) explicitly stated in a single sentence (i.e., ``right there''); ii) implicitly found only by integrating two or more sentences (i.e., ``think and search''), and iii) not in the text, forcing the readers to use their own knowledge (i.e., ``on my own'').  Their classification scheme is mostly designed to assess children's questions.  Graesser \cite{graesser1994question} investigated how to classify questions within the context of college-level mathematical tutoring sessions.  Questions are classified based on the length of the expected answers---short-answer questions (e.g. Verification, Quantification) require only a word or a phrase, while long-answer questions (e.g. Comparison, Interpretation) requires deeper reasoning and more elaborated answers.  In Gallagher and Ascher's framework \cite{gallagher1963preliminary}, convergent-thinking and divergent-thinking questions are equivalent to text-explicit and text-implicit questions of Raphael and Pearson \cite{raphael_increasing_1985}, respectively.


\begin{figure*}[t!]
\centering
\subfigure[Choosing Proposition]{
\fbox{\includegraphics[width=0.475\linewidth]{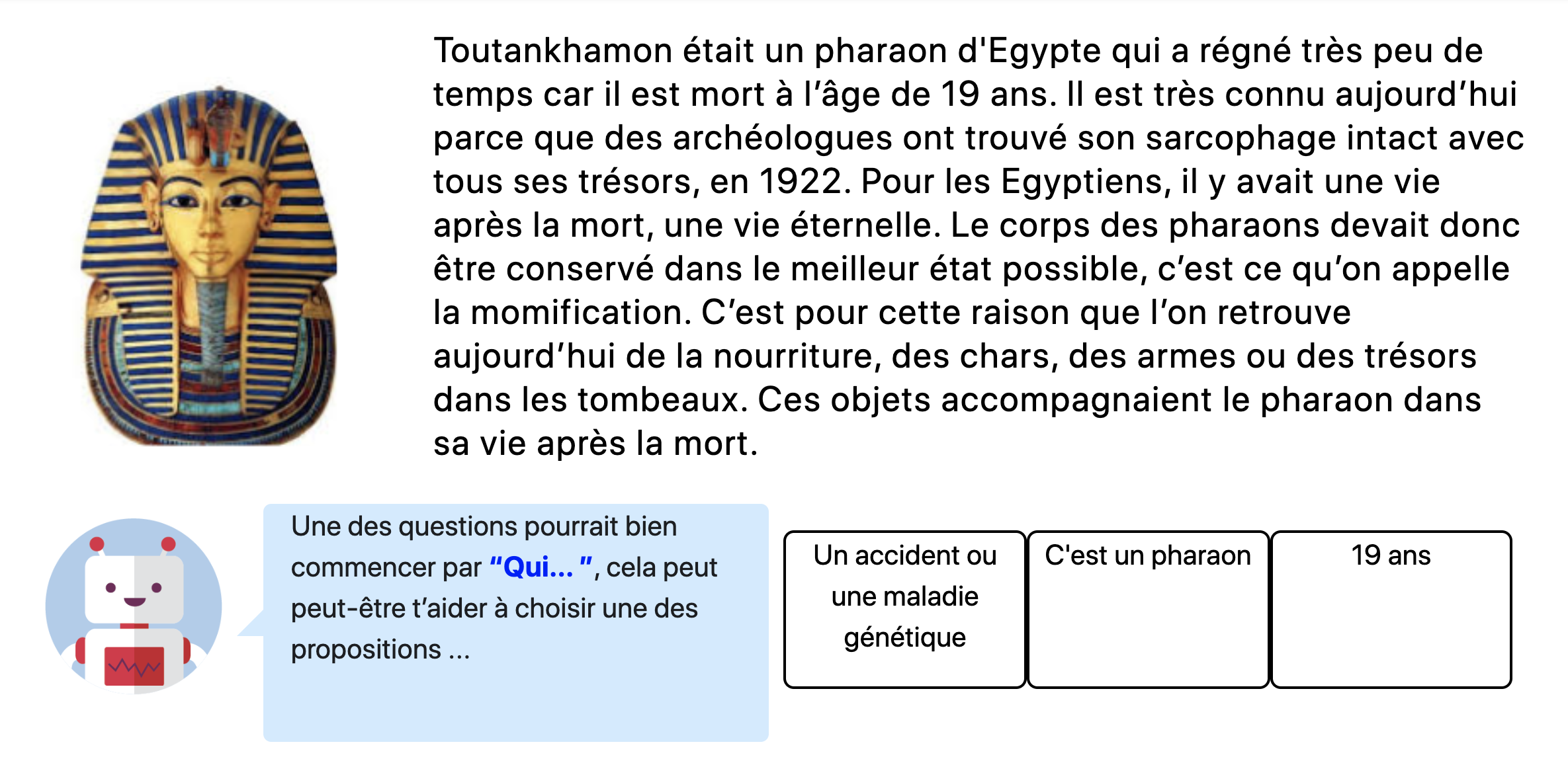}}
\label{fig:choose}
}
\subfigure[Generating Question]{
\fbox{\includegraphics[width=0.475\linewidth]{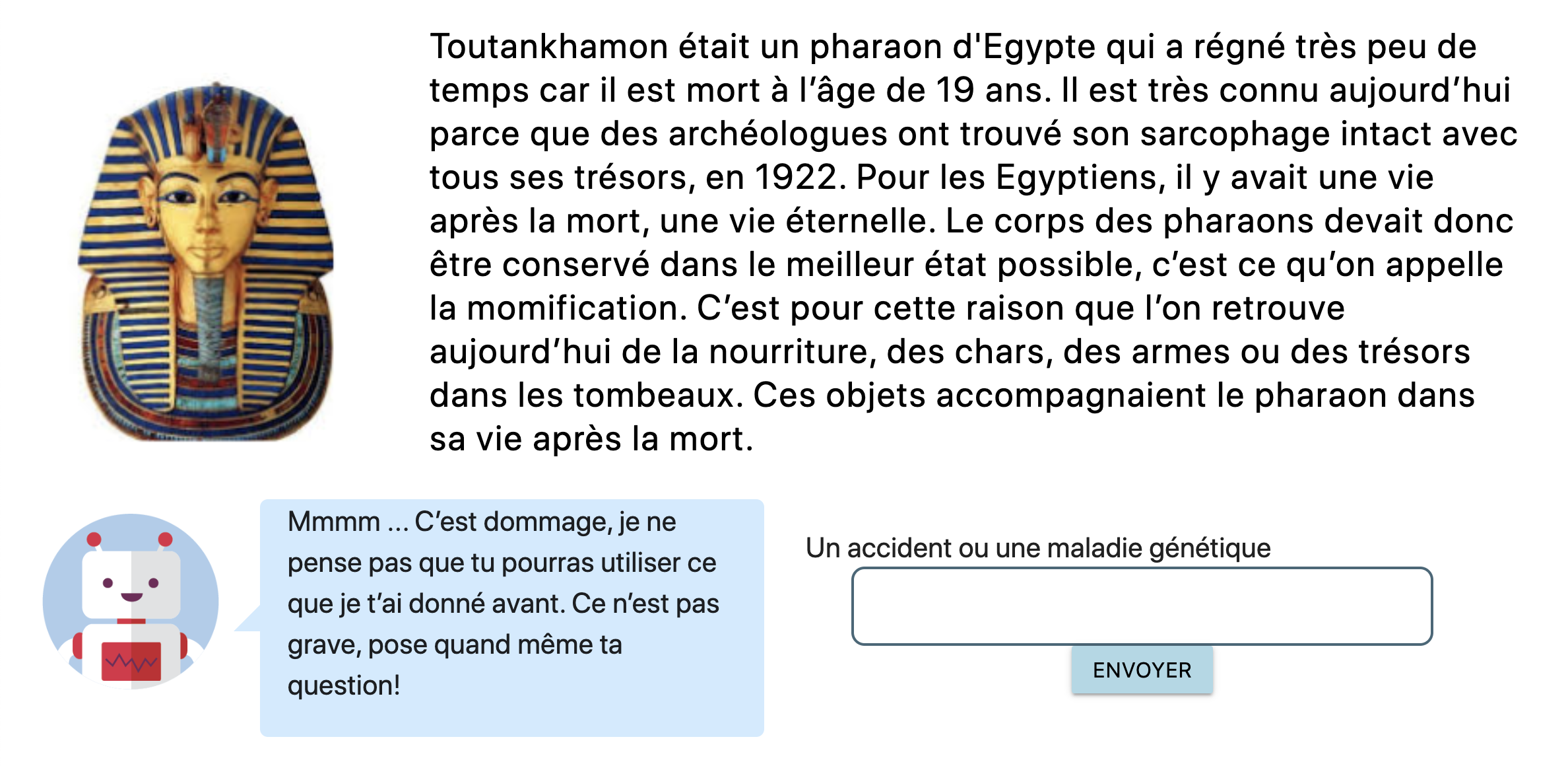}}
\label{fig:question}
}
\caption{Interface for choosing proposition (left) and generating question (right)}
\label{fig:interventions}
\end{figure*}

\subsection{Epistemic curiosity and Question-Asking}

Graesser et al. \cite{graesser1992mechanisms,graesser2008question} outlines four psychological needs that motivate the generation of a question: the need to monitor common ground to assess what other people know, the need to socially coordinate actions (e.g., ask for a request or an advice), the need to control conversation and attention, and important to our study---the need to correct a knowledge deficit in order to obtain a piece of missing information---a psychological need related to epistemic curiosity.  

Epistemic curiosity is a form of intrinsic motivation for acquiring knowledge, that can be both a state (i.e., a stable feeling or moment of interest or deprivation) and a trait (i.e., a general propensity to seek curiosity states) \cite{litman_nature_2005,litman_measurement_2004}. One mechanism for eliciting curiosity is through a {\it knowledge gap}.  When information is missing, or contradicts what one knows, a knowledge goal will arise, often leading to the generation of questions. The person is then made aware of the information needs, and motivated to formulate a question to obtain the missing knowledge \cite{ram_theory_1991}. 

Research has revealed a strong connection between curiosity and question asking abilities.  In Jirout and Klahr \cite{jirout_childrens_2011}, children were given a set of options to explore, where the options can be of different levels of uncertainty.  Children who prefer to resolve greater amounts of uncertainty are defined as being more curious, and those who prefer lower levels of uncertainty are defined as being less curious.  Their study found a positive relationship between children's curiosity and their range of question asking abilities, i.e., children who are more curious ask more questions, are better at using questions to solve simple problems, and are more skilled in discriminating between helpful and unhelpful questions.  In our work, we also investigate the role that curiosity trait might play in question asking.

\vspace{-0.15cm}
\subsection{Educational Applications for Enhancing Curiosity}


Some prior work has shown that although teachers asked most (i.e., 95\%) of the questions in the classroom \cite{graesser1994question,graesser1992mechanisms}, they are not always good role models---only 4\% of their questions are high-level questions \cite{dillon1988remedial}.  One possibility is to train teachers to ask questions effectively \cite{thompson1997training}.  Alternatively, one can develop educational applications and pedagogical/peer agents to elicit and reinforce students' curiosity \cite{Ceha2019,saerbeck_expressive_2010,clement2018,gordon2015can}, as curiosity and question asking are strongly linked.  Ceha et al. \cite{Ceha2019} introduced a game-playing robot that elicits emotional and behavioural curiosity from students by expressing its own curiosity about the topic.  In Kidlearn \cite{clement2018,roy15} or Kidbreath \cite{delmas2018conception}, algorithms based on the learning progress theory of curiosity were used to adapt the learning task to each child's abilities, and sequence the teaching to optimize for learning gains.  Results show that students learned faster and had higher intrinsic motivation scores, when their sequence of lessons was scheduled by curiosity-driven algorithms than traditional linear learning algorithms \cite{delmas2018conception}.  Together, these prior work suggest that designing educational technology to use curiosity to drive learning is a fruitful avenue to explore.  In our work, we take a practical approach to fostering curiosity by designing a pedagogical agent to facilitate the practice of question asking.  To our knowledge, there exists no application for this exact purpose.  
\subsection{The Use of Conversational Agents in Education}

There exist a number of powerful text- and voice- based conversational agents in education, for handling administrative and management tasks to foster productivity~\cite{goel2016jill}, providing emotional support and positive reinforcement, providing social presence~\cite{chou2003redefining, griol2013architecture}, scaffolding the practice of specific skills (e.g., language learning~\cite{fryer2006bots, shawar2007fostering}) and meta-cognitive strategies (e.g., reflect on learning process~\cite{grigoriadou2003dialogue}), supporting higher-level thinking skills (e.g., by encouraging explanations~\cite{Aleven1999}), and assessing students' learning (e.g., QuizBot~\cite{ruan19}).  

The educational benefits of conversational agents have been widely demonstrated.  Jill Watson~\cite{goel2016jill}, a virtual teaching assistant for the Georgia Tech course on Knowledge-Based Artificial Intelligence, demonstrates that conversational agents can help improve the efficiency of the administrative side of education, such as answering FAQs and posting announcements. AutoTutor \cite{Graesser2001} is a virtual avatar that helps students actively construct knowledge through conversations, and it has been shown to have a significant positive effects on student grades~\cite{Graesser2005}. In the learning of algebraic expressions, Heffernan and Koedinger~\cite{Heffernan2002,Heffernan2003} demonstrate that students who engaged in a dialog with Ms. Linquist, a virtual tutor, learned more.  QuizBot~\cite{ruan19} helps students learn factual knowledge and vocabulary.  Results show that students were able to recognize and recall over 20\% more correct answers compared to learning with flashcards.  In addition, students voluntarily spent significantly more time learning with Quizbot over flashcards. Finally, recent work has shown how voice-based conversational agents can be used to help young children read and learn~\cite{xu19}.  

Despite their demonstrated potential to impact education, conversational agents are limited by their ability to process natural language and recognize social and emotional cues, making them practical only for well-defined, narrow tasks~\cite{leonhardt2007using}. None of the prior work on conversational agents in education focused on question generation as a learning activity.




\section{Research Questions and Hypotheses}

Our main research question is how to design a virtual agent to help children improve their question-asking skills, in particular, their ability to generate divergent-thinking questions \cite{humphries2015beyond}.  The agent is embedded in a web application called the Curiosity Notebook \cite{law20}, a research infrastructure for studying teachable and pedagogical agents. On this platform, students can choose articles to read and are asked to generate questions related to the text.  Our study involves manipulating the agent and the interfaces to encourage students to generate different types of questions. Specifically, we hypothesize that such an agent can:

\begin{itemize}[noitemsep]
    \item help children construct more questions and questions that are of a high level of complexity (i.e., divergent-thinking questions) through the use of propositions and question starter prompts [H1].
    \item facilitate a question generation exercise that is enjoyable and motivating to do [H2].
    \item influence children's perception of the value of curiosity [H3].
\end{itemize}

To answer these research questions, we conducted an experiment with 95 elementary school children interacting with either an agent that promotes convergent thinking (i.e., the generation of memory-based or convergent-thinking questions) or one that promotes divergent thinking (i.e., the generation of curiosity-based, divergent-thinking questions). 

The convergent-thinking vs divergent-thinking question dichotomy comes from the well-known Question-Answer Relationship (QAR) classification, which stresses the relationship between a question, generated from a text reading, and its answer \cite{gallagher1963preliminary,raphael_increasing_1985,Wilen1991}. QAR describes two levels of questions: convergent-thinking questions for which the answers are in the text (e.g., low-level, factoid questions) and divergent-thinking questions for which the answers are not in the text, but that are text-elicited (e.g., ``What could happen if the main character went home instead of to the park?''). Table \ref{tab:baseline_dialogue} and \ref{tab:intervention_dialogue} shows how convergent-thinking and divergent-thinking questions can be defined by whether their corresponding answers can be found in the text.  For example, in Table \ref{tab:baseline_dialogue}, one of the convergent propositions is ``In Oslo", a phrase that is explicitly mentioned in the text as the location where the Nobel Peace Prize ceremony takes place.  On the other hand, one of the divergent propositions is ``6 different Nobel prizes", which is {\it not} found in the text.  The corresponding question---e.g., ``How many different kinds of Nobel prizes are there?"---is considered a divergent-thinking question because the answer is not found in the text.

Below, we outline our study design, including participant recruitment and experimental procedure, materials, data collection instruments and data cleaning process.

\section{Study Design}

\subsection{Experimental Procedure}

\begin{figure}[htbp!]
  \includegraphics[width=\columnwidth]{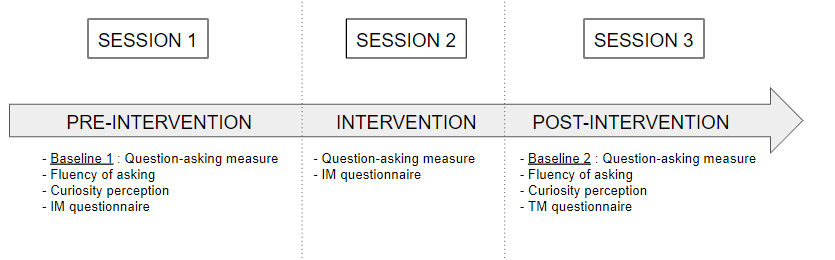}
  \caption{Study Timeline}
  \label{fig:timeline}
\end{figure}

The experiment is within-subject and involves a ABA design (Figure \ref{fig:timeline}), consisted of a pre-intervention baseline (A), intervention (B) and post-intervention baseline (A) sessions, where each session was held on a different day within the same week.  The length and number of sessions were limited to 3 days due to constraints imposed by the school schedule and ethics guidelines concerning studies with children.

\begin{table*}[t!]
    \centering
    \caption{Example showing the choices of propositions and student-agent dialogue in the Baseline condition}
    \small{
    \begin{tabular}{ p{5cm} p{2cm} p{4.5cm} p{4.5cm} }
    \hline
    ~ & ~ & ~ & ~\\
    \multicolumn{4}{p{17cm}}{{\bf Article}: The Nobel prizes reward people who advance the state of our knowledge and greatly benefit humankind. There are several kinds of Nobel prizes. In 2018, the Nobel prize for Physics was awarded to two researchers for inventing the world's most powerful laser beam. The Nobel Peace prize has been awarded to a doctor in the Congo, Denis Mukwege, and to an activist from Iraqi, Nadia Murad, for their fight against violence against women. The prize is handed out, along with money, during a special ceremony. This ceremony is held every year in Oslo, the capital of Norway.}\\\\
    {\bf Agent Speech} & ~ & {\bf Convergent Proposition} &  {\bf Divergent Proposition}\\
    Here are the responses to two questions.  Choose one of the propositions and try to find the question that it corresponds to.  Take your time to formulate! & ~ & Each year & 6 different Nobel prizes\\\\
    Two new propositions, choose one of them. I wonder what questions were asked ... & ~ & In Oslo & Some other examples are Literature, Chemistry and Medicine\\\\
    Last propositions for this text! Take your time to formulate your question! & ~ & Denis Mukwege & This laser can be used to heal the eyes\\\\
    \hline
    \end{tabular}}
    \label{tab:baseline_dialogue}
\end{table*}

\begin{table*}[t!]
    \centering
    \caption{Example showing the choices of propositions and student-agent dialogue in the Convergent and Divergent condition}
    \small{
    \begin{tabular}{ p{5cm} p{3.67cm} p{3.67cm} p{3.67cm}}
    \hline
     ~ & ~ & ~ & ~\\
    \multicolumn{4}{p{16cm}}{{\bf Article}: 6 volunteer scientists entered a very large white dome ... only to come out 358 days later! The goal was to live in the same conditions as astronauts who will be going to Mars, the red planet: dehydrated dishes, artificial lighting, the impossibility of going outside into the open air ... The goal is to be able to travel to the red planet in the future. The US Space Agency (NASA) conducted this test to see if it is possible to remain in good health and to work when being confined in a rocketship for such a long time.  Going to Mars is an expedition that would last at least two years!}\\\\
    \rowcolor{Gray}
    \multicolumn{4}{l}{\bf Convergent Condition}\\
    {\bf Agent Speech} & {\bf Convergent Proposition 1} & {\bf Convergent Proposition 2} & {\bf Divergent Proposition}\\
    One of the questions might start with ``Why...'', so that may help you choose one of the propositions ... & Being able to live on Mars & 358 days & Sports, Reading, Exercises\\\\
    One of the questions might start with ``What is ...'', so that may help you choose one of the propositions ... & The NASA & A 2 year trip  & Using powerful spaceships\\\\
    One of the questions might start with ``How many ...'', so that may help you choose one of the propositions ... & 6 scientists & Objective is to go to Mars & First trip will take place in 2030\\\\
    \rowcolor{Gray}
    \multicolumn{4}{l}{\bf Divergent Condition}\\
    {\bf Agent Speech} & {\bf Divergent Proposition 1} & {\bf Divergent Proposition 2} & {\bf Convergent Proposition}\\
    One of the questions might start with ``How...'' so that may help you choose one of the propositions ... & Sports, Reading, Exercices & There is no return trip & Being able to live on Mars\\\\
    One of the questions might start with ``How...'' so that may help you choose one of the propositions ... & Using powerful spaceships & We think we can live there & The NASA\\\\
    One of the questions might start with ``In which year...'', so that may help you choose one of the propositions ... & First trip will take place in 2030 & Nobody went to Mars yet & 6 scientists\\\\
    \hline
    \end{tabular}}
    \label{tab:intervention_dialogue}
\end{table*}

In each session, children were asked to generate questions for a set of short articles, one by one.  The system began by presenting children with a set of topics, and asked them to choose one topic (e.g., Olympics) from the list.  Upon choosing a topic, a short article related to the topic was displayed.  Children were asked to read the article, and indicate (by clicking on a button) when they are done reading.  After reading, children were presented with a set of propositions (Figure \ref{fig:choose}); the number of propositions depends on the session phase (baseline vs intervention), as will be described later.  A proposition is an answer to a question that can either be found in the text (convergent proposition), or cannot be found in the text but that is related to the content (divergent proposition).  Children were asked to choose one of the propositions to generate a question from (Figure \ref{fig:question}).  In other words, students had only the answers and they had to guess what questions would lead to such answers. For each article, children repeated this process of generating questions based on a choice of proposition three times; after that, there was a 4th {\it free question} round where children could generate any questions they wanted about the text, without being given any propositions.  When the 4 questioning rounds were finished for that article, children proceeded to choose another topic, read another piece of text, and generate questions.  

Children were not restricted in terms of time; they completed the experiment at their own pace.  Children interacted with the application on a tablet; they were summoned one by one to a quiet room to participate in the experiment, so that they are not distracted by the rest of the class. 

In the {\bf baseline} (i.e., A) sessions, students were asked to process 3 articles. For each article, the agent presents 2 propositions to the children based on which to generate a question; one of the propositions represents an answer to a convergent-thinking question, and the other to a divergent-thinking question. Table \ref{tab:baseline_dialogue} provides an example of a baseline round, including the article, the choices of propositions, and what the agent said when asking children to choose a proposition. As the example shows, the agent was passive---it did not help or influence children to choose one or the other proposition.  When presenting the propositions, the agent simply said, ``Here are the responses to two questions.  Choose one of the propositions and try to find the question that it corresponds to.  Take your time to formulate.''  

There were two baseline sessions, conducted before (baseline 1) and after (baseline 2) the intervention session. The two baseline manipulations were exactly the same, with only the articles being different. In each baseline session, children asked 12 questions in total, 3 based on propositions and 1 free question for each of the 3 articles.

In the {\bf Intervention} (i.e., B) session, students were asked to process 6 articles.  Similar to the baseline, for each article, they did 3 rounds of proposition-guided questioning plus a free question; but this time, they were given 3 propositions to choose from in each questioning round instead of two.  Students were divided into two intervention condition groups.  In the {\texttt Convergent} condition, two of the propositions were answers that would lead to convergent-thinking questions, while the third proposition was one that would lead to a divergent-thinking question.   Likewise, in the {\texttt Divergent} condition, two of the three propositions were directed towards divergent-thinking questions, with the third directing towards a convergent-thinking question.  An additional feature of the intervention session is that the agent will attempt to {\it prompt} the children to choose a particular proposition by giving them a {\it question starter}.  For example, while the children are choosing a proposition, the agent would say ``One of the questions might start with {\it When...}.  It might help you choose one of the propositions...''. In the Divergent condition, the agent would attempt to encourage divergent thinking by giving a question starter that would lead to a divergent-thinking question, and likewise for the Convergent condition.  




Table \ref{tab:intervention_dialogue} provides an example of an intervention round and highlights these differences between the Convergent and Divergent conditions. Only the choices of propositions and question starter prompts were different; the article was the same.  With this manipulation, our goal is to influence children to generate a certain type of questions (convergent-thinking or divergent-thinking), and investigate whether there is any immediate or longer term influence on their question generation behaviour post-intervention (i.e., in baseline 2).  In total, each child asked 24 questions---18 questions guided by propositions and prompts, and 6 free questions---during the intervention session. 

Over the entire experiment, each child generated 36 questions guided by propositions---9 questions in each baseline session, and 18 questions in the intervention session. 


\subsection{Design Rationale}
In developing the pedagogical agent, one of the design choices we made is to provide propositions and question starter prompts to make the process of question generation easier for children. The rationale behind this decision is two fold.  First, Humphries and Ness \cite{humphries2015beyond} found that children, without external aids, have great difficulty generating divergent-thinking questions, and proposed a set of verbal tools that help students construct questions for which the answers are not directly in the text. Inspired by this work, we provide simple vs complex question starters as verbal tools to help children generate questions. Second, Graesser and Person \cite{graesser1994question} argues that children fail to generate divergent-thinking questions because they are unable to identify their own knowledge gap. Suggesting answers to children helps to reveal their knowledge gap, while simultaneously serving as a form of gamification---since the divergent propositions are not in the text, the novelty of the propositions and the challenge of finding the right questions serve as design elements for keeping children curious and engaged.

\subsection{Materials}


The 12 articles used in the experiment were selected from a variety of children literature, including a magazine (i.e., Images doc), book/encyclopedia (i.e., La grande imagerie) and website (i.e., 1 jour 1 actu), related to subjects in science and history that are likely to interest children.  They were edited in such a way that children can quickly read and understand them---the edited articles all have 6 sentences and a mean of 18 $\pm$ 7 words per sentence. For each article, the authors generated the propositions based on the two levels of questions (convergent-thinking and divergent-thinking) specified in Gallagher and Ascher's classification scheme~\cite{gallagher1963preliminary}, which is known to be particularly useful for categorizing children's questions \cite{humphries2015beyond}.  To facilitate the replication of this experiment, our dataset, which includes articles, propositions and prompts, is available at  \url{https://doi.org/10.7910/DVN/JKD52Y}.

\subsection{Participants}

We recruited 95 5th grade students belonging to 4 classes from an elementary school, aged between 10 and 12 years old.  After baseline 1, participants were randomly assigned to 2 intervention condition groups (Convergent or Divergent). As shown in Table \ref{tab:demographics} and confirmed via t- or $\chi^{2}$ tests, the two groups did not differ in term of demographics (age, gender) as well as other profile measures.  

\begin{table}[h!]
    \centering
    \caption{Mean demographics and profile measure for the Convergent vs Divergent condition}
    \small{
    \begin{tabular}{p{2.5cm} p{1.2cm} p{1.2cm} p{1.5cm}}
    \hline
         ~ & Convergent  & Divergent  & t/$\chi^2$\\
         ~ & (n=38) & (n=34) & p values\\
         \hline
          age & 10.8  & 10.9  & t = -1.50\\
          (years) & ($\pm$0.27) & ($\pm$0.46) &  p=.139\\
          \hline
          gender  & 21M &  21M & $\chi^2$ = 0.0764  \\
          (M/F) & 21F & 19F & p=.782\\
           \hline
          
          reading ability  & 167 & 160  & t = 0.83 \\
          (\# read word/min) &  ($\pm$37.27) & ($\pm$37.59) & p= 410\\
           \hline
          verbal understanding & 18,88  & 18,50  & t = -0.018\\
           (max=42) & ($\pm$6,14) & ($\pm$5,01) & p=.986\\
           \hline
          curiosity trait  & 28,84  & 29,06  & t =-0.223 \\
          (max=40) & ($\pm$3,78) & ($\pm$4,46) & p=.824\\
           \hline
          device use freq. & 3.13  & 3.18  & t = -0.294\\
           (1 to 4 score) & ($\pm$0.71) & ($\pm$0.63) & p= .770\\
         \hline
    \end{tabular}}
    \label{tab:demographics}
\end{table}

We removed the data of 2 children, who had learning disabilities and were participating in the experiment with the help of a school assistant.  We removed students who were not present during all three sessions, students whose verbal outputs are not deemed usable (e.g., they entered mostly gibberish), and students whose majority of generated questions were incomplete, unrelated to the chosen proposition or not well-formulated. In total, we removed 23 children's data; this leaves 72 participants for our final analysis, composed of 33 boys and 39 girls.  The data cleaning procedure is further discussed below.


\subsection{Data Collection Instruments}

\subsubsection{Session Data} 
From the Curiosity Notebook, we collected the proposition that children chose and the question that they generated in each questioning round.  From this data, we produced a count of how many convergent-thinking versus divergent-thinking questions were generated. We refer to this count as the question-asking score (max score=18 in the intervention session; max score=9 in each baseline session).


\subsubsection{Profile} 
This includes information about the child's age, gender, 
verbal understanding (WISC-IV subscale \cite{wechsler03}), reading abilities (E.L.F.E test \cite{lequette2008elfe}), and curiosity trait \cite{litman_measurement_2004}.  This information was collected at the beginning of the first session, and the curiosity trait questionnaire was given to the parents to complete before the experiment.

\subsubsection{Intrinsic Motivation and Type of Motivation} 
The {\it Intrinsic Motivation Scale} (IMS, max score = 40)  measures children's motivation in using the application \cite{cordova1996intrinsic}.   This was administered before and after the intervention, i.e., during the pre- and post-intervention baseline sessions.  The {\it Type of Motivation} (TM) questionnaire \cite{vallerand1992academic}, which is a superset of the IMS, was used to assess the type of motivation elicited by the intervention. It is divided in 3 subscales, which include Amotivation (AM, max score = 3), Extrinsic Motivation (EM, max score = 9) and Intrinsic Motivation (IM, max score = 9).  The TM questionnaire was administered at the end of the post-intervention baseline session.  Together, these scales allow us to evaluate the type and level of motivation behind the children interacting with the agent.



\subsubsection{Fluency of Question Asking}  
The {\it Fluency of Question Asking} test measures the number of questions that children can freely generate about a piece of text without any external aids.  In this test, children were told to read a short piece of text (specifically, about ants), and to ask as many questions as they can within 1 minute.   In order to assess whether our intervention has any effects on children's fluency of question asking, this test was administered both at the beginning of the first session, as well as at the end of the last session. 

\subsubsection{Perception of the Value of Curiosity} 
The Children's Images of and Attitudes Towards Curiosity (CIAC) questionnaire measures elementary school children's perception of and attitude towards curiosity \cite{post_development_2019}. The 24-item questionnaire consists of two components.  The first component consists of a 2-item {\it Image of Curiosity} scale, which measures how much students relate social matters to curiosity, and a 5-item {\it Epistemic Image of Curiosity} scale, which measures how much children associate epistemic questions to curiosity.  The second component consists of the {\it Attitude towards Epistemic Curiosity} scale covering: Personal inclination (PR, 4 items) which measures how children perceive the benefit of question asking in class and their degree of enjoyment doing it; Social Relevance (SR, 3 items), which assesses the extent to which children see curiosity having any social relevance; Negative Opinion (NO, 3 items), which evaluates whether children perceive the act of question-asking in a negative way; Fear of Classmates Negative Judgement (FCNJ, 3 items), which evaluates children's level of fear of being judged by other people in the classroom when asking questions; and finally, Self-Efficiency (SE, 4 items), which measures how children perceive their own skills in asking questions at school. To test whether our interventions have any effects on children's perception of the value of curiosity, these surveys were administered both at the beginning of the first session, as well as at the end of the last session.

All questionnaires have been converted to 4-point Likert scales to ensure a homogeneous analysis of data, except for the TM questionnaire which contains yes/no questions only \cite{vallerand1992academic}. This is based on prior research \cite{kulas2009middle} showing that a Likert scale with an even number of items typically forces children to make a decision between positive and negative answers, whereas with an uneven number of items, children are more likely to choose the middle item as the answer.



\subsection{Data Cleaning}


Since the question asking scores are based on questions generated from propositions, we processed the data to ensure that children entered well-formulated questions.  

\begin{table}[h!]
    \centering
    \caption{Percentage of well-formulated questions for each session}
    \small{
    \begin{tabular}{p{1.2cm} p{1.8cm} p{1.8cm} p{1.8cm}}
    \hline
         ~ & Baseline 1 & Intervention & Baseline 2\\
         \hline
          Convergent & 90.74 ($\pm$12.48) & 89.74 ($\pm$12.56) & 90.46 ($\pm$11.23)\\
          \hline
         Divergent & 93.07 ($\pm$15.59) & 89.89 ($\pm$15.43) & 93.5 ($\pm$11.21)\\
         \hline
    \end{tabular}}
    \label{tab:wellformulated}
\end{table}

Our analyses are based on counting divergent-thinking and convergent-thinking questions. In each round, if children chose a convergent/divergent proposition, the question they generate is considered a convergent-thinking/divergent-thinking question, unless it is deemed incorrect. A question is considered correct if the chosen proposition is the correct answer to that question. 
As an example, for the proposition {\it 200 million years ago}, an acceptable question would be ``When were dinosaurs alive?''  We also accepted questions like ``Dinosaurs were alive, when?'' or ``Dinosaurs lived, when?'', but not ``Were dinosaurs alive 200 millions years ago?'' to which the appropriate answer is yes or no. 

Two raters coded 10\% of the questions, and considered 79\% and 86\% of the questions asked to be correct (i.e., the answer to the question matches the proposition) respectively.  The inter-rater reliability is 82.5\% overall, 82\% for convergent-thinking questions and 83\% for divergent-thinking questions.    Table \ref{tab:wellformulated} shows the number of well-formulated questions that we retained for analysis, by session and by intervention condition.

\section{Results}

\subsection{Question-asking performance during intervention}

\begin{figure}[h!]
  \includegraphics[width=\columnwidth]{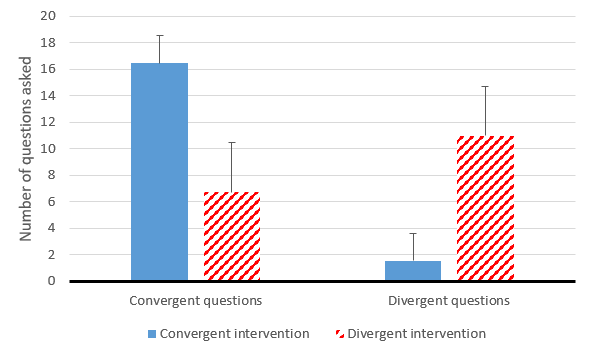}
  \caption{Number of convergent-thinking vs divergent-thinking questions asked in the intervention session by intervention condition}
  \label{fig:condition_results}
\end{figure}

There was a significant difference in the number of divergent-thinking questions asked during the intervention session between the two conditions (t(670) = -13.6 , p < 0.001, $\eta^2$= -3.22), as shown in Figure \ref{fig:condition_results}. The Divergent condition elicited a higher percentage (61\%) and average number of divergent-thinking questions (m = 11.10; SD = 3.74) compared to the Convergent condition (6\%; m = 6.85; SD = 2.08). We also observed that children more frequently chose to ask questions that were prompted by the pedagogical agent; 73\% of questions asked in the Divergent condition and 70\% of the questions asked in the Convergent condition were generated using question starters provided by the agent. 


In addition, the results suggest that divergent-thinking questions are much more difficult to generate than convergent-thinking questions.  Even with prompting, children in the Divergent condition still opted to ask convergent-thinking questions almost 40\% of the time; conversely, the incentive provided by the agent in the Convergent condition resulted in 94\% convergent-thinking questions.

\subsection{Mediating effect of curiosity trait}

In analyzing the mediating effects of curiosity trait, we conducted an ANCOVA with curiosity trait as co-variate.  Results show a slight trend of curiosity trait score mediating the type of questions asked (F(1,69) = 3.13, p = 0.08, $\eta^2$ = 0.011). 

\begin{figure}[h!]
  \includegraphics[width=\columnwidth]{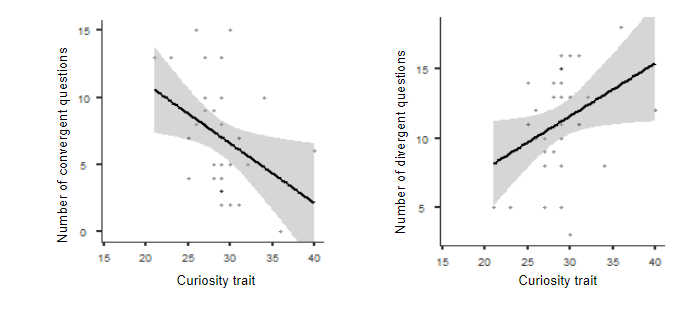}
  \caption{Correlation between the number of convergent-thinking (left) or divergent-thinking (right) questions asked and the curiosity trait of students in the Divergent condition}
  \label{fig:curiosity_trait}
\end{figure}

For this reason, we performed a correlation analysis between curiosity trait and each type of question, in both conditions. There is no significant correlation in the Convergent condition. However, as Figure \ref{fig:curiosity_trait} shows, in the Divergent condition, the number of divergent-thinking questions is positively correlated to curiosity trait (r = 0.358, p = 0.038) and the number of convergent-thinking questions is negatively correlated to curiosity trait (r = -0.390, p = 0.023). It means that the more curious a student is (by trait), the more divergent-thinking questions and the fewer convergent-thinking questions they will ask.

\subsection{Intervention Effect: Pre-Post Measures}

Lastly, we were interested in the influence of the intervention on question asking behaviour (i.e., in terms of the number of divergent-thinking questions asked and fluency of question asking), motivation and perception of the value of curiosity, as measured by pre- and post- measures.

\subsubsection{Performance of Question Asking}

Mixed ANOVA revealed a significant difference in the number of divergent-thinking questions asked bewteen the pre-intervention and post-intervention baselines (F(1,70) = 13.76, p < 0.01, $\eta^2$ = 0.074). 

\begin{figure}[h!]
  \includegraphics[width=\columnwidth]{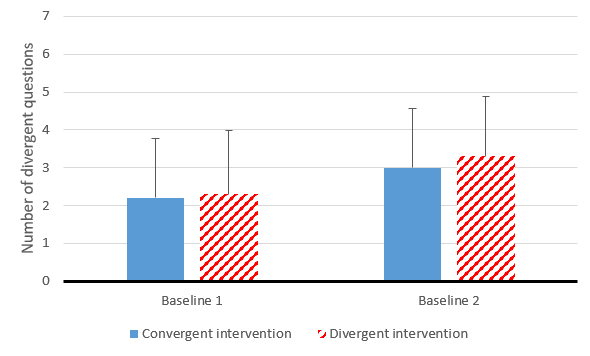}
  \caption{Number of divergent-thinking questions asked during Baseline 1 vs Baseline 2 by intervention condition}
  \label{fig:typeofquestionprepost}
\end{figure}

As Figure 5 shows, students asked more divergent-thinking questions in post-intervention baseline (m = 3.14; SD = 1.57) than in the pre-intervention baseline (m = 2.25 ; SD = 1.62). However, the two-way effect (intervention condition $\times$ baseline) did not reach significance (p > 0.60). Taken together, these results indicate that all children have improved their capability to ask divergent-thinking questions irrespective of which intervention condition (Convergent vs. Divergent) they were assigned to. With 18 trials of question asking training, both interventions yielded the same benefit in terms of encouraging children to ask more divergent-thinking questions. 

\subsubsection{Motivational Measures}

In terms of intrinsic motivation (IM), there was no significant difference between the two baseline sessions (p = 0.535) or between the intervention groups (p = 0.228). The scores are relatively high (m = 28.7, out of 40) and remain stables throughout the experiment.

\begin{figure}[t!]
  \includegraphics[width=\columnwidth]{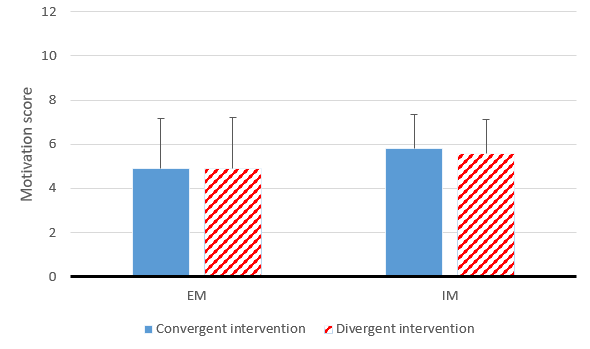}
  \caption{Motivation score of students by type of motivation (intrinsic vs extrinsic) and intervention condition}
  \label{fig:convergent}
\end{figure}

In terms of the type of motivation, statistical analyses revealed a difference between extrinsic and intrinsic motivation scores (F(1,70) = 12.62, p < 0.001, $\eta^2$ = 0.046). Students seem to be more intrinsically motivated (m = 5.72 $\pm$ 6) than extrinsically motivated (m = 4.88 $\pm$ 5), irrespective of intervention condition (F(1,70) = 0.0265, p = 0.871).  There is no interaction between intervention condition and type of motivation (F(1,70) = 0.109, p = 0.773).

\subsubsection{Fluency of Question Asking}

Children asked more questions in post-intervention baseline (m = 9.0 $\pm$ 2.7) than in pre-intervention baseline (m= 6.8 $\pm$ 2.5); this result is significant according to Mixed ANOVA analysis (F(1,70)= 47.12, p < 0.001, $\eta^2$ = 0.147). There were no significant differences in the number of questions asked between the two intervention groups (F(1,70) = 1.13, p = 0.290) and no interaction between the intervention condition and baselines (F(1,70) = 1.23, p = 0.271).

\begin{figure}[h!]
  \includegraphics[width=\columnwidth]{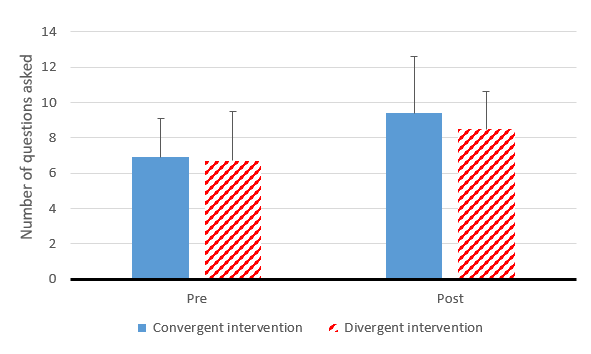}
  \caption{Number of questions asked during the pre vs post-intervention fluency of question asking test, by intervention condition}
  \label{fig:convergent}
\end{figure}

\subsubsection{Curiosity Perception}


In terms of the image of curiosity, mixed ANOVA shows that there are no significant differences between the baselines  (F(1,70) = 0.067, p = 0.931) or between intervention conditions (F(1,70) = 0.061, p = 0.804), and there is no two-way interaction (F(1,70) = 0.027, p = 0.870). Students had nearly the same perception or image of curiosity at the end of the experiment as the beginning.

\begin{table}[h!]
    \centering
    \caption{Pre and post mean image of curiosity score by intervention}
    \small{
    \begin{tabular}{c c c}
    \hline
         ~ & Pre (/4) & Post (/4)\\
         \hline
          Convergent Condition & 2.15 ($\pm$ 0.57) & 2.25 ($\pm$ 0.48)\\
          Divergent Condition & 2.69 ($\pm$ 0.56) & 2.63 ($\pm$ 0.57)\\
         \hline
    \end{tabular}}
    \label{tab:imageOfCuriosityScore}
\end{table}


In terms of attitude toward epistemic curiosity, Mixed ANOVA revealed no significant differences between the baselines  (F(1,70) = 1.98, p = 0.164) or between intervention conditions (F(1,70) = 0.132, p = 0.718), and there is no two-way interaction (F(1,70) = 0.003, p = 0.956). Students had nearly the same attitude toward epistemic curiosity at the end of the experiment as the beginning.

\begin{table}[h!]
    \centering
    \caption{Pre and post mean attitude toward epistemic curiosity score by intervention condition}
    \small{
    \begin{tabular}{c c c}
    \hline
         ~ & Pre (/4) & Post (/4)\\
         \hline
          Convergent Condition & 2.30 ($\pm$ 0.48) & 2.27 ($\pm$ 0.51)\\
          Divergent Condition & 2.34 ($\pm$ 0.46) & 2.29 ($\pm$ 0.52)\\
         \hline
    \end{tabular}}
    \label{tab:meanAttitude}
\end{table}


\section{Discussion}

This work explores how pedagogical agents can be used to improve the ability of children to ask more questions and questions of higher levels of complexity. Specifically, we proposed and evaluated an intervention that can steer children to ask more questions of a particular type.  Our main results show that both interventions led to higher fluency of question asking and a significant increase in the number of divergent-thinking questions asked, when comparing the post-intervention to the pre-intervention results.  This is counter to our initial assumption, that the Divergent condition will result in a bigger improvement of question-asking skills than the Convergent condition.  One simple explanation is that the repeated practice of asking questions (which is common to both interventions) led children to be more comfortable with generating questions, as well as questions that require curiosity and a greater amount of information seeking. This confirms existing theories \cite{engel2011children} which established a strong connection between children's mastery of question generation mechanics and their curiosity-related behaviors. The results also resonate with curiosity studies which show the positive cyclical effects of question asking~\cite{Browne2007,Chin2008,graesser1994question}---the more question you ask, the more curious you become, leading to even more questions.  Our work went one step further by demonstrating that pushing children to ask more questions can lead them to ask more complex ones.


Other more nuanced conclusions can be drawn from the study. First, the findings provide some evidence that our intervention was effective: the agent was able to successfully influence children to ask either more convergent-thinking or divergent-thinking questions through the use of propositions and question starter prompts. In the Divergent condition, our agent led children to generate 61\% divergent-thinking questions, which is noteworthy in light of the previous findings of Humphries and Ness \cite{humphries2015beyond} reporting that children would ask 93\% convergent-thinking questions when no incentives were given. As observed, 73\% of divergent-thinking questions asked in the Divergent condition were those prompted by question-starters.  Without question starters prompting divergent-thinking questions (as is the case in the Convergent condition), only 6\% of the questions that children asked were divergent-thinking. Thus, it is possible {\it and} extremely beneficial to foster question asking skills with the help of a pedagogical agent, who provides hints and models higher-level question asking behaviour \cite{humphries2015beyond}.

Second, results show that the curiosity trait of children was positively correlated with the number of divergent-thinking questions asked in the Divergent condition.  In other words, children who are generally curious benefited more from the Divergent intervention.  
This connection between curiosity trait and ability to ask divergent-thinking questions is consistent with prior work \cite{loewenstein_psychology_1994,jirout_childrens_2011}, which suggests that curious children tend to focus on implicit information in text, seek new knowledge more often, and are better at discriminating between helpful and unhelpful questions than less curious children. Therefore, the Divergent intervention---of having an agent encourage divergent-thinking questions---is particularly fruitful for the highly curious children, whereas the Convergent intervention is beneficial for all the children.

As a final argument in favor of the use of educational agents to facilitate question asking training, across both interventions, the intrinsic motivation score remained high and superior to extrinsic motivation. This shows that children generally find the question-asking learning activity to be enjoyable and motivating.

\subsection{Limitations}
On the flip side, our results also showed that the intervention effects were short-lived---there is a drastic decrease of divergent-thinking questions performance for the Divergent group, from 61\% in the intervention session to 36\% in the post-intervention baseline.  One possible reason is the sudden removal of support by the agent when transitioning from the intervention session to the post-intervention baseline session, which made the task of question asking more difficult.  This discrepancy in children's performance between agent-supported and self-supported question asking mirrors a well-known learning phenomenon---the maintenance of under-developed skills requires support from the environment, whereas well-developed skills that can be self-sustained regardless of the circumstances \cite{sauzeon2012developmental}. In other words, the 18 trials of question asking training may simply be inadequate to consolidate the learning of divergent-thinking question asking behaviors.   


The intervention duration can also explain why no positive change was observed in children's perception of and attitude towards curiosity. This is not surprising in light of the nature (i.e., digital interaction) and duration (i.e., 3 days) of the performed intervention. Indeed, it is well documented in social attitude literature that reliable and sustainable attitudinal changes often require longer interventions and/or with more realistic social interaction with teacher or between children [10]. Moreover, as reported by Post and Walma van der Molen \cite{post_development_2019}, negative age-related attitudinal changes toward epistemic curiosity is common during the transitional period between Grade 5 and Grade 6. Researchers have found that children's perception about the learning value and use of epistemic curiosity may decline as they progress through primary school.

\subsection{Long Term Benefits}
Overall, the present results suggest that an agent can be used to enhance question-asking skills in children, irrespective of whether the Convergent or Divergent intervention was used.
Previous research found the generation of divergent-thinking questions~\cite{claxton07} to be an effective exercise for fostering higher-level thinking skills in young students---motivating them to voice inquisitive ideas, make creative connections between knowledge domains, and seek alternative solutions to problems~\cite{engel15,heywood12,osborne2014}---and for enhancing their reading comprehension skills \cite{ness_simple_2017}. The convergent/divergent question distinction was designed to assess children's questions \cite{humphries2015beyond}, and therefore can be also used to educate teachers on how to recognize questioning behaviour in the classroom and how to use effectively use questioning themselves as a pedagogical strategy~\cite{leven81}.  Therefore, as a classroom tool, our pedagogical agent has the potential to benefit children and teachers alike. 




\section{Conclusion}


In this work, we developed a pedagogical agent that encourages children to ask divergent-thinking questions, a more complex form of questions that is associated with curiosity. We conducted a study with 95 fifth grade students, who interacted with an agent that encourages either convergent-thinking or divergent-thinking questions. Results showed that both interventions increased the number of divergent-thinking questions and the fluency of question asking, while they did not significantly alter children's perception of curiosity despite their high intrinsic motivation scores.  In addition, children's curiosity trait has a mediating effect on question asking under the agent that promoted divergent-thinking, suggesting that question-asking interventions must be personalized to each student based on their tendency to be curious.



Despite encouraging results, we observed that the question asking exercise was still too difficult for some children. In future work, we could develop a more guided approach or an adaptive strategy where we prompt for questions of different levels of complexity, in order to avoid a cognitive load \cite{sweller_cognitive_1994}. We hypothesize that different types of text (e.g interesting vs boring, familiar vs unfamiliar, text containing complete vs partial information) may influence children's epistemic curiosity and the way they generate questions \cite{duffy1989models}; as such, future work can explore new types of materials and how they mediate the effects of agent-facilitated question asking. Currently, questions were coded manually after the experiment and not processed automatically during the interaction with the agent. In future work, one can automate processing by parsing question starters (e.g., what, when, how) from children's questions, or by creating a topic model to describe the questions and the main text to ensure some correspondence between the two. Finally, we can investigate short and long term interventions on question asking and assess their effects on learning outcomes.



\section{Acknowledgments}

This work was funded by HFSP Grant RGP0018/2016, Idex/University of Bordeaux and NSERC Discovery Grant RGPIN-2015-0454.  We would also like to thank Cecile Mazon (Ph.D. Cognitive Science) for her help in the data cleaning and coding, Didier Roy (Researcher in Education Science) for his support in the recruitment of students, and all the participants in this studies, including students, parents and teachers.

\balance{}

\bibliographystyle{SIGCHI-Reference-Format}
\bibliography{main,extra}


\begin{thebibliography}{00}


\ifx \showCODEN    \undefined \def \showCODEN     #1{\unskip}     \fi
\ifx \showDOI      \undefined \def \showDOI       #1{{\tt DOI:}\penalty0{#1}\ }
  \fi
\ifx \showISBNx    \undefined \def \showISBNx     #1{\unskip}     \fi
\ifx \showISBNxiii \undefined \def \showISBNxiii  #1{\unskip}     \fi
\ifx \showISSN     \undefined \def \showISSN      #1{\unskip}     \fi
\ifx \showLCCN     \undefined \def \showLCCN      #1{\unskip}     \fi
\ifx \shownote     \undefined \def \shownote      #1{#1}          \fi
\ifx \showarticletitle \undefined \def \showarticletitle #1{#1}   \fi
\ifx \showURL      \undefined \def \showURL       #1{#1}          \fi

\bibitem{Aleven1999}
{Vincent Aleven}, {Kenneth~R Koedinger}, {and} {Karen Cross}. 1999.
\newblock \showarticletitle{{Tutoring Answer Explanation Fosters Learning with
  Understanding}}. In {\em Proceedings of the 9th International Conference on
  Artificial Intelligence in Education} {\em (AIED '99)}. IOS Press, Amsterdam,
  199--206.
\newblock


\bibitem{brown_instructing_1984}
{Ann Brown}, {Annemarie Palincsar}, {and} {Bonnie Armbruster}. 2013.
\newblock \showarticletitle{{Instructing Comprehension-Fostering Activities in
  Interactive Learning Situations}}.
\newblock In {\em Theoretical Models and Processes of Reading}. International
  Reading Association, Newark, DE, 657--689.
\newblock
\showDOI{%
\url{http://dx.doi.org/10.1598/0710.27}}


\bibitem{Browne2007}
{Neil~M. Browne} {and} {Stuart~M. Keeley}. 2007.
\newblock {\em {Asking The Right Questions: A Guide to Critical Thinking}}.
\newblock Prentice-Hall, Englewood Cliffs, N.J.
\newblock


\bibitem{Ceha2019}
{Jessy Ceha}, {Joslin Goh}, {Corina McDonald}, {Dana Kuli{\'{c}}}, {Edith Law},
  {Nalin Chhibber}, {and} {Pierre-Yves Oudeyer}. 2019.
\newblock \showarticletitle{{Expression of Curiosity in Social Robots: Design,
  Perception, and Effects on Behaviour}}. In {\em CHI Conference on Human
  Factors in Computing Systems} {\em (CHI 2019)}. ACM, New York, NY, USA,
  1--12.
\newblock
\showDOI{%
\url{http://dx.doi.org/10.1145/3290605.3300636}}


\bibitem{Chin2008}
{Christine Chin} {and} {Jonathan Osborne}. 2008.
\newblock \showarticletitle{{Students' Questions: A Potential Resource for
  Teaching and Learning Science}}.
\newblock {\em Studies in Science Education\/} {44}, 1 (2008), 1--39.
\newblock
\showDOI{%
\url{http://dx.doi.org/10.1080/03057260701828101}}


\bibitem{chou2003redefining}
{Chih-Yueh Chou}, {Tak-Wai Chan}, {and} {Chi-Jen Lin}. 2003.
\newblock \showarticletitle{{Redefining the Learning Companion: the Past,
  Present, and Future of Educational Agents}}.
\newblock {\em Computers \& Education\/} {40}, 3 (2003), 255--269.
\newblock
\showDOI{%
\url{http://dx.doi.org/10.1016/S0360-1315(02)00130-6}}


\bibitem{claxton07}
{Guy Claxton}. 2007.
\newblock \showarticletitle{{Expanding Young People's Capacity to Learn}}.
\newblock {\em British Journal of Educational Studies\/} {55}, 2 (2007),
  115--134.
\newblock
\showDOI{%
\url{http://dx.doi.org/10.1111/j.1467-8527.2007.00369.x}}


\bibitem{clement2018}
{Benjamin Cl{\'e}ment}. 2018.
\newblock {\em {Adaptive Personalization of Pedagogical Sequences using Machine
  Learning}}.
\newblock Ph.D. Dissertation. University of Bordeaux.
\newblock


\bibitem{collins2005discriminative}
{Michael Collins} {and} {Terry Koo}. 2005.
\newblock \showarticletitle{{Discriminative Reranking for Natural Language
  Parsing}}.
\newblock {\em Computational Linguistics\/} {31}, 1 (2005), 25--70.
\newblock
\showDOI{%
\url{http://dx.doi.org/10.1162/0891201053630273}}


\bibitem{cordova1996intrinsic}
{Diana~I. Cordova} {and} {Mark~R. Lepper}. 1996.
\newblock \showarticletitle{{Intrinsic Motivation and the Process of Learning:
  Beneficial Effects of Contextualization, Personalization, and Choice}}.
\newblock {\em Journal of educational psychology\/} {88}, 4 (1996), 715--730.
\newblock
\showDOI{%
\url{http://dx.doi.org/10.1037/0022-0663.88.4.715}}


\bibitem{Davey1986}
{Beth Davey} {and} {Susan McBride}. 1986.
\newblock \showarticletitle{{Generating Self-Questions after Reading: A
  Comprehension Assist for Elementary Students}}.
\newblock {\em Journal of Educational Research\/} {80}, 1 (1986), 43--46.
\newblock
\showISSN{19400675}
\showDOI{%
\url{http://dx.doi.org/10.1080/00220671.1986.10885720}}


\bibitem{delmas2018conception}
{Alexandra~A. Delmas}. 2018.
\newblock {\em {Conception et Validation d'un Jeu d'Auto-Apprentissage de
  Connaissances sur l'Asthme pour le Jeune Enfant: R{\^o}le de la Motivation
  Intrins{\`e}que}}.
\newblock Ph.D. Dissertation. University of Bordeaux.
\newblock


\bibitem{dillon1988remedial}
{James~T. Dillon}. 1988.
\newblock \showarticletitle{{The Remedial Status of Student Questioning}}.
\newblock {\em Journal of Curriculum Studies\/} {20}, 3 (1988), 197--210.
\newblock
\showDOI{%
\url{http://dx.doi.org/10.1080/0022027880200301}}


\bibitem{duffy1989models}
{Thomas~M. Duffy}, {Lorraine Higgins}, {Brad Mehlenbacher}, {Cynthia Cochran},
  {David Wallace}, {Charles Hill}, {Diane Haugen}, {Margaret McCaffrey},
  {Rebecca Burnett}, {Sarah Sloane}, {and} {Suzanne Smith}. 1989.
\newblock \showarticletitle{{Models for the Design of Instructional Text}}.
\newblock {\em Reading Research Quarterly\/} {24}, 4 (1989), 434--457.
\newblock
\showDOI{%
\url{http://dx.doi.org/10.2307/747606}}


\bibitem{durlak2011impact}
{Joseph~A. Durlak}, {Roger~P. Weissberg}, {Allison~B. Dymnicki}, {Rebecca~D.
  Taylor}, {and} {Kriston~B. Schellinger}. 2011.
\newblock \showarticletitle{{The Impact of Enhancing Students' Social and
  Emotional Learning: A Meta-Analysis of School-Based Universal
  Interventions}}.
\newblock {\em {Child Development}\/} {82}, 1 (2011), 405--432.
\newblock
\showDOI{%
\url{http://dx.doi.org/10.1111/j.1467-8624.2010.01564.x}}


\bibitem{engel2011children}
{Susan Engel}. 2011.
\newblock \showarticletitle{{Children's Need to Know: Curiosity in Schools}}.
\newblock {\em Harvard educational review\/} {81}, 4 (2011), 625--645.
\newblock
\showDOI{%
\url{http://dx.doi.org/10.17763/haer.81.4.h054131316473115}}


\bibitem{engel15}
{Susan Engel}. 2015.
\newblock {\em {The Hungry Mind: The Origins of Curiosity in Childhood}}.
\newblock Hard University Press, Cambridge, MA.
\newblock


\bibitem{fryer2006bots}
{Luke Fryer} {and} {Rollo Carpenter}. 2006.
\newblock \showarticletitle{{Bots as Language Learning Tools}}.
\newblock {\em {Language Learning \& Technology}\/} {10}, 3 (2006), 8--14.
\newblock
\showDOI{%
\url{http://dx.doi.org/10125/44068}}


\bibitem{gallagher1963preliminary}
{James~J. Gallagher} {and} {Mary~Jane Aschner}. 1963.
\newblock \showarticletitle{{A Preliminary Report on Analyses of Classroom
  Interaction}}.
\newblock {\em Merrill-Palmer Quarterly of Behavior and Development\/} {9}, 3
  (1963), 183--194.
\newblock


\bibitem{goel2016jill}
{Ashok~K. Goel} {and} {Lalith Polepeddi}. 2016.
\newblock {\em {Jill Watson: A Virtual Teaching Assistant for Online
  Education}}.
\newblock {T}echnical {R}eport. Georgia Institute of Technology.
\newblock


\bibitem{gordon2015can}
{Goren Gordon}, {Cynthia Breazeal}, {and} {Susan Engel}. 2015.
\newblock \showarticletitle{{Can Children Catch Curiosity from a Social
  Robot?}}. In {\em {Proceedings of ACM/IEEE International Conference on
  Human-Robot Interaction}} {\em (HRI '15)}. ACM, New York, NY, USA, 91--98.
\newblock
\showDOI{%
\url{http://dx.doi.org/10.1145/2696454.2696469}}


\bibitem{graesser2008question}
{Art Graesser}, {Vasile Rus}, {and} {Zhiqiang Cai}. 2008.
\newblock \showarticletitle{Question classification schemes}. In {\em
  Proceedings of the Workshop on Question Generation}. 10--17.
\newblock


\bibitem{Graesser2005}
{Arthur~C. Graesser}, {Patrick Chipman}, {Brian~C. Haynes}, {and} {Andrew
  Olney}. 2005.
\newblock \showarticletitle{{Auto Tutor: An Intelligent Tutoring System with
  Mixed-Initiative Dialogue}}.
\newblock {\em IEEE Transactions on Education\/} {48}, 4 (2005), 612--618.
\newblock
\showISSN{00189359}
\showDOI{%
\url{http://dx.doi.org/10.1109/TE.2005.856149}}


\bibitem{graesser1992mechanisms}
{Arthur~C. Graesser}, {Natalie Person}, {and} {John Huber}. 1992.
\newblock \showarticletitle{{Mechanisms that Generate Questions}}.
\newblock In {\em {Questions and Information Systems}}, {T.~W. Lauer}, {Peacock
  E.}, {and} {A.~C. Graesser} (Eds.). Lawrence Erlbaum Associates, Mahwah, NJ,
  167--187.
\newblock


\bibitem{graesser1994question}
{Arthur~C. Graesser} {and} {Natalie~K. Person}. 1994.
\newblock \showarticletitle{{Question Asking during Tutoring}}.
\newblock {\em {American Educational Research Journal}\/} {31}, 1 (1994),
  104--137.
\newblock
\showDOI{%
\url{http://dx.doi.org/10.3102/00028312031001104}}


\bibitem{Graesser2001}
{Arthur~C. Graesser}, {Kurt VanLehn}, {Carolyn~P. Rose}, {Pamela~W. Jordan},
  {and} {Derek Harter}. 2001.
\newblock \showarticletitle{{Intelligent Tutoring Systems with Conversational
  Dialogue}}.
\newblock {\em {AI Magazine}\/} {22}, 4 (Dec. 2001), 39--51.
\newblock
\showDOI{%
\url{http://dx.doi.org/10.1609/aimag.v22i4.1591}}


\bibitem{grigoriadou2003dialogue}
{M Grigoriadou}, {G Tsaganou}, {and} {Th Cavoura}. 2003.
\newblock \showarticletitle{{Dialogue-based Reflective System for Historical
  Text Comprehension}}. In {\em {Proceedings of Workshop on Learner Modelling
  for Reflection at Artificial Intelligence in Education, International
  Conference on Artificial Intelligence in Education}}. 238--247.
\newblock


\bibitem{griol2013architecture}
{David Griol} {and} {Zoraida Callejas}. 2013.
\newblock \showarticletitle{{An Architecture to Develop Multimodal Educative
  Applications with Chatbots}}.
\newblock {\em {International Journal of Advanced Robotic Systems}\/} {10}, 175
  (2013), 1--15.
\newblock
\showDOI{%
\url{http://dx.doi.org/10.5772/55791}}


\bibitem{Heffernan2003}
{N~T Heffernan}. 2003.
\newblock \showarticletitle{{Web-based Evaluations Showing Both Cognitive and
  Motivational Benefits of the Ms. Lindquist Tutor}}.
\newblock In {\em {Artificial Intelligence in Education: Shaping the Future of
  Learning through Intelligent Technologies}}, {Ulrich Hoppe}, {Felisa
  Verdejo}, {and} {Judy Kay} (Eds.). IOS Press, 115--122.
\newblock


\bibitem{Heffernan2002}
{Neil~T. Heffernan} {and} {Kenneth~R. Koedinger}. 2002.
\newblock \showarticletitle{{An Intelligent Tutoring System Incorporating a
  Model of an Experienced Human Tutor}}. In {\em {Intelligent Tutoring Systems.
  Lecture Notes in Computer Science, vol 2363.}}, {Stefano~A. Cerri}, {Guy
  Gouard{\`e}res}, {and} {F{\`a}bio Paragua{\c{c}}u} (Eds.). Springer, {Berlin,
  Heidelberg}, 596--608.
\newblock
\showDOI{%
\url{http://dx.doi.org/10.1007/3-540-47987-2_61}}


\bibitem{heilman2010good}
{Michael Heilman} {and} {Noah~A. Smith}. 2010.
\newblock \showarticletitle{Good Question! Statistical Ranking for Question
  Generation}. In {\em Human Language Technologies: Annual Conference of the
  North American Chapter of the Association for Computational Linguistics} {\em
  (HLT '10)}. ACL, Stroudsburg, PA, USA, 609--617.
\newblock
\showISBNx{1-932432-65-5}
\showURL{%
\url{http://dl.acm.org/citation.cfm?id=1857999.1858085}}


\bibitem{heywood12}
{David Heywood}, {Joan Parker}, {and} {Nina Jolley}. 2012.
\newblock \showarticletitle{{Pre-service Teachers' Shifting Perceptions of
  Cross-curricular Practice: The Impact of School Experience in Mediating
  Professional Insight}}.
\newblock {\em International Journal of Educational Research\/}  {55} (12
  2012), 89--99.
\newblock
\showDOI{%
\url{http://dx.doi.org/10.1016/j.ijer.2012.07.003}}


\bibitem{hohmann1995educating}
{Mary Hohmann}, {David~P. Weikart}, {and} {Ann~S. Epstein}. 1995.
\newblock {\em {Educating Young Children: Active Learning Practices for
  Preschool and Child Care Programs}}.
\newblock High/Scope Press, Ypsilanti, MI.
\newblock


\bibitem{humphries2015beyond}
{Jean Humphries} {and} {Molly Ness}. 2015.
\newblock \showarticletitle{{Beyond Who, What, Where, When, Why, and How:
  Preparing Students to Generate Questions in the Age of Common Core
  Standards}}.
\newblock {\em Journal of Research in Childhood Education\/} {29}, 4 (2015),
  551--564.
\newblock
\showDOI{%
\url{http://dx.doi.org/10.1080/02568543.2015.1073199}}


\bibitem{jirout_childrens_2011}
{Jamie Jirout} {and} {David Klahr}. 2011.
\newblock \showarticletitle{{Children's Question Asking and Curiosity: A
  Training Study Conference}}. In {\em Proceedings of the Society for Research
  on Educational Effectiveness Conference}. 1--4.
\newblock


\bibitem{king_effects_1989}
{Alison King}. 1989.
\newblock \showarticletitle{{Effects of Self-Questioning Training on College
  Students' Comprehension of Lectures}}.
\newblock {\em Contemporary Educational Psychology\/} {14}, 4 (1989), 366--381.
\newblock
\showDOI{%
\url{http://dx.doi.org/10.1016/0361-476X(89)90022-2}}


\bibitem{king_autonomy_1994}
{Alison King}. 1994.
\newblock \showarticletitle{{Autonomy and Question Asking: The Role of Personal
  Control in Guided Student-Generated Questioning}}.
\newblock {\em Learning and Individual Differences\/} {6}, 2 (1994), 163--185.
\newblock
\showDOI{%
\url{http://dx.doi.org/10.1016/1041-6080(94)90008-6}}


\bibitem{kulas2009middle}
{John~T. Kulas} {and} {Alicia~A. Stachowski}. 2009.
\newblock \showarticletitle{{Middle Category Endorsement in Odd-numbered Likert
  Response Scales: Associated Item Characteristics, Cognitive Demands, and
  Preferred Meanings}}.
\newblock {\em Journal of Research in Personality\/} {43}, 3 (2009), 489--493.
\newblock
\showDOI{%
\url{http://dx.doi.org/10.1016/j.jrp.2008.12.005}}


\bibitem{law20}
{Edith Law}, {Parastoo~Baghaei Ravari}, {Nalin Chhibber}, {Dana Kulic},
  {Stephanie Lin}, {Kevin~D. Pantasdo}, {Jessy Ceha}, {Sangho Suh}, {and}
  {Nicole Dillen}. 2020.
\newblock \showarticletitle{Curiosity Notebook: A Platform for Learning by
  Teaching Conversational Agents}. In {\em Submission to Extended Abstracts of
  the SIGCHI Conference on Human Factors in Computing Systems} {\em (CHI '20)}.
  1--8.
\newblock


\bibitem{leonhardt2007using}
{Michelle~Denise Leonhardt}, {Liane Tarouco}, {Rosa~Maria Vicari},
  {Elder~Rizzon Santos}, {and} {Michele dos~Santos da Silva}. 2007.
\newblock \showarticletitle{{Using Chatbots for Network Management Training
  through Problem-based Oriented Education}}. In {\em Proceedings of IEEE
  International Conference on Advanced Learning Technologies} {\em (ICALT
  '07)}. IEEE, 845--847.
\newblock
\showDOI{%
\url{http://dx.doi.org/10.1109/ICALT.2007.275}}


\bibitem{lequette2008elfe}
{C Lequette}, {G Pouget}, {and} {M Zorman}. 2008.
\newblock ELFE. {\'E}valuation de la Lecture en Fluence.
\newblock   (2008).
\newblock
\newblock
\shownote{\url{http://www.cognisciences.com/accueil/outils/article/e-l-fe-evaluation-de-la-lecture-en-fluence}.}


\bibitem{leven81}
{Tamar Leven} {and} {Ruth Long}. 1981.
\newblock {\em {Effective Instruction}}.
\newblock Association for Supervision and Curriculum Development, Washington,
  DC.
\newblock


\bibitem{litman_nature_2005}
{Jordan~A. Litman}, {Robert~P. Collins}, {and} {Charles~D. Spielberger}. 2005.
\newblock \showarticletitle{{The nature and measurement of sensory curiosity}}.
\newblock {\em Personality and Individual Differences\/} {39}, 6 (2005),
  1123--1133.
\newblock
\showDOI{%
\url{http://dx.doi.org/10.1016/j.paid.2005.05.001}}


\bibitem{litman_measurement_2004}
{Jordan~A. Litman} {and} {Tiffany~L. Jimerson}. 2004.
\newblock \showarticletitle{{The Measurement of Curiosity as a Feeling of
  Deprivation}}.
\newblock {\em Journal of Personality Assessment\/} {82}, 2 (2004), 147--157.
\newblock
\showDOI{%
\url{http://dx.doi.org/10.1207/s15327752jpa8202_3}}


\bibitem{loewenstein_psychology_1994}
{George Loewenstein}. 1994.
\newblock \showarticletitle{{The Psychology of Curiosity: A Review and
  Reinterpretation}}.
\newblock {\em Psychological Bulletin\/} {116}, 1 (1994), 75--98.
\newblock
\showDOI{%
\url{http://dx.doi.org/10.1037/0033-2909.116.1.75}}


\bibitem{marx_effects_1999}
{Alexandra Marx}, {Urs Fuhrer}, {and} {Terry Hartig}. 1999.
\newblock \showarticletitle{{Effects of Classroom Seating Arrangements on
  Children's Question-Asking}}.
\newblock {\em Learning Environments Research\/} {2}, 3 (1999), 249--263.
\newblock
\showDOI{%
\url{http://dx.doi.org/10.1023/A:1009901922191}}


\bibitem{ness_simple_2017}
{Molly Ness}. 2017.
\newblock \showarticletitle{{Simple Texts, Complex Questions: Helping Young
  Children Generate Questions}}.
\newblock {\em Reading Improvement\/} {54}, 1 (2017), 1--5.
\newblock


\bibitem{osborne2014}
{Jonathan Osborne}. 2014.
\newblock \showarticletitle{{Teaching Scientific Practices: Meeting the
  Challenge of Change}}.
\newblock {\em Journal of Science Teacher Education\/} {25}, 2 (2014),
  177--196.
\newblock
\showDOI{%
\url{http://dx.doi.org/10.1007/s10972-014-9384-1}}


\bibitem{post_development_2019}
{Tim Post} {and} {Juliette H.~Walma van~der Molen}. 2019.
\newblock \showarticletitle{{Development and Validation of a Questionnaire to
  Measure Primary School Children's Images of and Attitudes towards Curiosity
  (the CIAC Questionnaire)}}.
\newblock {\em Motivation and Emotion\/} {43}, 1 (2019), 159--178.
\newblock
\showDOI{%
\url{http://dx.doi.org/10.1007/s11031-018-9728-9}}


\bibitem{ram_theory_1991}
{Ashwin Ram}. 1991.
\newblock \showarticletitle{{A Theory of Questions and Question Asking}}.
\newblock {\em Journal of the Learning Sciences\/} {1}, 3-4 (1991), 273--318.
\newblock
\showDOI{%
\url{http://dx.doi.org/10.1080/10508406.1991.9671973}}


\bibitem{raphael_increasing_1985}
{Taffy~E. Raphael} {and} {P.~David Pearson}. 1985.
\newblock \showarticletitle{{Increasing Students' Awareness of Sources of
  Information for Answering Questions}}.
\newblock {\em American Educational Research Journal\/} {22}, 2 (1985),
  217--235.
\newblock
\showDOI{%
\url{http://dx.doi.org/10.3102/00028312022002217}}


\bibitem{rosenshine_teaching_1996}
{Barak Rosenshine}, {Carla Meister}, {and} {Saul Chapman}. 1996.
\newblock \showarticletitle{{Teaching Students to Generate Questions: A Review
  of the Intervention Studies}}.
\newblock {\em Review of Educational Research\/} {66}, 2 (1996), 181--221.
\newblock
\showDOI{%
\url{http://dx.doi.org/10.3102/00346543066002181}}


\bibitem{roy15}
{Didier Roy}. 2015.
\newblock {\em {Personnalisation Automatique des Parcours d'Apprentissage dans
  les Syst{\`e}mes Tuteurs Intelligents}}.
\newblock Research report. {Inria Bordeaux Sud-Ouest}.
\newblock
\showURL{%
\url{https://hal.inria.fr/hal-01144515}}


\bibitem{ruan19}
{Sherry Ruan}, {Liwei Jiang}, {Justin Xu}, {Bryce Joe-Kun Tham}, {Zhengneng
  Qiu}, {Yeshuang Zhu}, {Elizabeth~L. Murnane}, {Emma Brunskill}, {and}
  {James~A. Landay}. 2019.
\newblock \showarticletitle{QuizBot: A Dialogue-based Adaptive Learning System
  for Factual Knowledge}. In {\em Proceedings of the SIGCHI Conference on Human
  Factors in Computing Systems} {\em (CHI '19)}. ACM, New York, NY, USA, 1--13.
\newblock
\showDOI{%
\url{http://dx.doi.org/10.1145/3290605.3300587}}


\bibitem{saerbeck_expressive_2010}
{Martin Saerbeck}, {Tom Schut}, {Christoph Bartneck}, {and} {Maddy~D. Janse}.
  2010.
\newblock \showarticletitle{{Expressive Robots in Education: Varying the Degree
  of Social Supportive Behavior of a Robotic Tutor}}. In {\em Proceedings of
  the SIGCHI Conference on Human Factors in Computing Systems} {\em (CHI '10)}.
  ACM, New York, NY, USA, 1613--1622.
\newblock
\showISBNx{978-1-60558-929-9}
\showDOI{%
\url{http://dx.doi.org/10.1145/1753326.1753567}}


\bibitem{sauzeon2012developmental}
{H{\'e}l{\`e}ne Sauz{\'e}on}, {Marie D{\'e}jos}, {Philippe Lestage}, {Prashant
  Arvind~Pala}, {and} {Bernard N'Kaoua}. 2012.
\newblock \showarticletitle{{Developmental Differences in Explicit and Implicit
  Conceptual Memory Tests: A Processing View Account}}.
\newblock {\em Child Neuropsychology\/} {18}, 1 (2012), 23--49.
\newblock
\showDOI{%
\url{http://dx.doi.org/10.1080/09297049.2011.557652}}


\bibitem{shawar2007fostering}
{Bayan~Abu Shawar} {and} {Eric Atwell}. 2007.
\newblock \showarticletitle{{Fostering language learner autonomy through
  adaptive conversation tutors}}. In {\em Proceedings of the Corpus Linguistics
  Conference}. 1--8.
\newblock


\bibitem{sweller_cognitive_1994}
{John Sweller}. 1994.
\newblock \showarticletitle{{Cognitive Load Theory, Learning Difficulty, and
  Instructional Design}}.
\newblock {\em Learning and Instruction\/} {4}, 4 (1994), 295--312.
\newblock
\showDOI{%
\url{http://dx.doi.org/10.1016/0959-4752(94)90003-5}}


\bibitem{thompson1997training}
{Geoff Thompson}. 1997.
\newblock \showarticletitle{{Training Teachers to Ask Questions}}.
\newblock {\em ELT journal\/} {51}, 2 (1997), 99--105.
\newblock
\showDOI{%
\url{http://dx.doi.org/10.1093/elt/51.2.99}}


\bibitem{vallerand1992academic}
{Robert~J Vallerand}, {Luc~G Pelletier}, {Marc~R Blais}, {Nathalie~M Briere},
  {Caroline Senecal}, {and} {Evelyne~F Vallieres}. 1992.
\newblock \showarticletitle{{The Academic Motivation Scale: A Measure of
  Intrinsic, Extrinsic, and Amotivation in Education}}.
\newblock {\em Educational and Psychological Measurement\/} {52}, 4 (1992),
  1003--1017.
\newblock
\showDOI{%
\url{http://dx.doi.org/10.1177/0013164492052004025}}


\bibitem{wechsler03}
{David Wechsler}. 2003.
\newblock {\em {Wechsler Intelligence Scale for Children Fourth Edition:
  Technical and Interpretive Manual}}.
\newblock Psychological Corporation, San Antonio, TX.
\newblock


\bibitem{Wilen1991}
{W. Wilen}. 1991.
\newblock {\em {Questioning Skills for Teachers: What Research Says to the
  Teacher (Third Edition)}}.
\newblock National Education Association, Washington, DC.
\newblock


\bibitem{xu19}
{Ying Xu} {and} {Mark Warschauer}. 2019.
\newblock \showarticletitle{{Young Children's Reading and Learning with
  Conversational Agents}}. In {\em Extended Abstracts of the SIGCHI Conference
  on Human Factors in Computing Systems} {\em (CHI EA '19)}. ACM, New York, NY,
  USA, 1--8.
\newblock
\showDOI{%
\url{http://dx.doi.org/10.1145/3290607.3299035}}


\end{thebibliography}

\end{document}